\documentclass[reqno]{article}
\usepackage{amssymb}
\usepackage{amsmath}
\usepackage{graphicx}
\usepackage{amsfonts}

\input{tcilatex}
\begin{document}

\title{\textbf{STABILITY ANALYSIS OF FRACTIONAL DIFFERENTIAL EQUATIONS WITH
UNKNOWN PARAMETERS}}
\author{\textbf{Mehmet Emir Koksal } \\
\\
Ondokuz Mayis University, Department of Mathematics, 55139 Samsun,
Turkey}
\date{}
\maketitle

\begin{quote}
\textbf{{\Large {Abstract }}}

In this paper, the stability of fractional differential equations (FDEs)
with unknown parameters is studied. FDEs bring many advantages to model the
physical systems in nature or man-made systems in industry. In fact, real
objects are generally fractional and fractional calculus has gained
popularity in modelling physical and engineering systems in the last few
decades in parallel to advancement of high speed computers. Using the
graphical based D-decomposition method, we investigate the parametric
stability analysis of FDEs without complicated mathematical analysis. To
achieve this, stability boundaries are obtained firstly, and then the
stability region set depending on the unknown parameters is found. The
applicability of the presented method is shown considering some benchmark
equations which are often used to verify the results of a new method.
Simulation examples shown that the method is simple and give reliable
stability results.

\end{quote}

\section{INTRODUCTION}

\setcounter{section}{1}

\bigskip Fractional differential equations \cite[2-5]{1} are a
generalization of classical integer order differential equations through the
application of fractional calculus \cite[7-10]{6} which has been developed
by pure mathematicians firstly since after half of the 19th century though
engineers and physicist found applications of fractional calculus for
various concepts 100 years later \cite{11}. As a field of mathematical
analysis, fractional calculus studies the possibility of taking real or
complex number powers of differential operators. It may be considered an old
branch of mathematical analysis, but it is a novel topic yet \cite{12}.
Especially, fractional calculus has gained a great deal of popularity in
modelling some physical and engineering systems as well as fractal phenomena
in the last few decades \cite[14-19]{13}. In fact, many systems in the real
world are now better characterized by FDEs and analysed by numerical
techniques developed for solving differential equations involving
non-integer order derivatives \cite{20}. FDEs are also known as
extraordinary differential equations.

Continuing technological developments have required new methods in basic
sciences, especially in mathematics for analysis and design of physical
systems and their control tools. These methods which are easily implemented
with the advancement of high speed computers aimed to better and better
characterization, design tools and control performance of modern
technological products of engineering systems of developing civilization.
These development which had covered only static system models involving
geometry and algebra until 1965, had started using dynamical models
involving differential and integral calculus since 1965; and now have been
accelerating since the 1960's with the fractional order modelling involving
FDEs which have gained force and dare with high speed computers \cite[22]{21}%
. Hence FDEs have become a powerful tool in studying, designing and control
of physical systems and engineering products of the present world and it
still constitutes a research area in need.

Stability is one of the most important objects in the analysis and design of
dynamical systems. If the differential equation of a system has not a stable
property, the system may burn out, disintegrate or saturate when a signal is
applied \cite{23}. Therefore an unstable system is useless in practice and
needs a stabilization process via an additional control element mostly \cite[%
25]{24}. If a system has unknown parameters, the stability analysis is
called as parametric stability analysis. It is more difficult than the
classical stability analysis which has simple analysis methods such as
Routh-Hurwitz method, Nyquist stability theorem, etc. In the literature,
there are some methods on the stability of FDEs with uncertain parameters
\cite{26, 27, 28}. The uncertainty in these studies is considered by a
certain interval of the unknown parameters. However, to consider the unknown
parameters' values in a whole parametric interval from zero to infinity is
more useful than a certain little interval, which is the subject of this
paper.

Motivated by the need of stability analysis for FDEs with unknown
parameters, we suggest in this paper an efficient graphical based stability
analysis using the D-decomposition method \cite[30]{29}. The D-decomposition
method provides a powerful and simple stability work environment to the
analyst. This method is based on a conformal mapping from frequency domain
to parametric domain of unknown parameters. With this mapping, the imaginary
axis which is the stability bound of complex s-plane converts to three types
of stability boundaries, which are named as real, complex and infinite root
boundaries, in the parametric space. These boundaries give us the stability
regions which are important tools including useful stability knowledge. The
algorithm presented in this paper has a reliable result which is illustrated
by several examples, and is practically useful in the computerized analysis
of FDEs having unknown parameters.

The paper is organized as follows: The basic principles of FDEs are
revisited in the next section. The stability concept for FDEs is explained
in Section 3. A derivation of the stability boundary formulae for the
stability regions of fractional differential equations with unknown
parameters is given in Section 4. The next section illustrates the
effectiveness of the stability analysis proposed with three simulation
examples. Finally, Section 6 gives some concluding remarks.

\section{PRINCIPLES OF FRACTIONAL DIFFERENTIAL EQUATIONS}

\setcounter{section}{2}

In the most general case, FDEs are expressed by the following form \cite{31}:%
\begin{equation}
F\left( t,y(t),_{0}D_{t}^{\alpha _{1}}y(t),\cdots ,_{0}D_{t}^{\alpha
_{n}}y(t)\right) =G\left( t,u(t),_{0}D_{t}^{\beta _{1}}u(t),\cdots
,_{0}D_{t}^{\beta _{m}}u(t)\right)  \tag{1}  \label{1}
\end{equation}%
where $F$ and $G$ are fractional differential functions, $\alpha _{i}$ $%
(i=1\sim n)$ and $\beta _{k}$ $(k=1\sim m)$ are positive real numbers such
that $0<\alpha _{1}<\alpha _{2}<\cdots <\alpha _{n},~0<\beta _{1}<\beta
_{2}<\cdots <\beta _{m}$ and $m<n.$ $_{\alpha }D_{t}^{\gamma }$ is
fractional order derivative and integral operator and it is defined as
follows \cite{32}:
\begin{equation}
_{\alpha }D_{t}^{\gamma }=\left\{
\begin{array}{c}
d^{\gamma }/dt\gamma \text{ \ \ }\Re (\gamma )>0, \\
1\text{ \ \ \ \ \ \ \ \ \ \ \ }\Re (\gamma )=0, \\
\int_{a}^{t}(d\tau )^{-\gamma }\text{ }\Re (\gamma )<0.%
\end{array}%
\right.  \tag{2}  \label{2}
\end{equation}%
Here $\gamma $ is fractional order, $\Re (\gamma )$ is the real part of
fractional order and $a$ is a constant coefficient related with initial
conditions. Commonly, $t$ is an independent variable representing time, $%
u(t) $ is input exiting function and $y(t)$ is the output response function
of a dynamical system. There are various definitions for fractional
derivative. Riemann-Liouville, Gr\"{u}nwald-Letnikov, Caputo and
Mittag-Leffler are the well-known and common definitions among them. (For a
more detail of these definitions, the reader can see \cite{3, 32}).

One of the most common types of FDEs is linear time-invariant fractional
differential equations%
\begin{equation}
\sum_{i=0}^{n}a_{i\text{ }0}D_{t}^{^{\alpha _{i}}}y(t)=\sum_{k=0}^{m}b_{k%
\text{ }0}D_{t}^{^{\beta _{k}}}u(t),\text{ }\alpha _{n}\neq 0,\text{ }\alpha
_{0}=0,  \tag{3}  \label{3}
\end{equation}%
where $a_{i\text{ }}$and $b_{k\text{ }}$ are real numbers. Eq. (\ref{3}) is
also called as non-commensurate order FDE. As a special case, fractional
orders $a_{i\text{ }}$and $b_{k\text{ }}$ may be multiple of same real
number $\varepsilon $ like $a_{i}=i\varepsilon $ and $b_{k}=k\varepsilon $.
In this case, Eq. (\ref{3}) is named by commensurate order FDE \cite[33, 34]%
{3}.

The Laplace transform method is commonly used in engineering systems and
their analysis. According to Gr\"{u}nwald-Letnikov definition, Laplace
transform of differential operator $_{a}D_{t}^{\gamma }$ is given by%
\begin{equation}
L\left\{ _{0}D_{t}^{\gamma }f(t)\right\} =s^{\gamma }F(s),  \tag{4}
\label{4}
\end{equation}%
where $s$ is the Laplace operator. For the differential equation in Eq. (\ref%
{3}), the transfer function giving the input-output expression of a system
with zero initial conditions is given by

\begin{equation}
G(s)=~~~\frac{Y(s)}{U(s)}=~~\frac{\sum_{i=0}^{m}b_{i}s^{\beta _{i}}}{%
a_{o}+\sum_{i=1}^{n}a_{i}s^{\alpha _{i}}}=\frac{N(s)}{D(s)}.  \tag{5}
\label{5}
\end{equation}%
In this equation, $U(s)$ is the Laplace transform of the exciting function $%
u(t)$; similarly $Y(s)$ is that of the response $y(t)$. $N(s)$ and $D(s)$
are the numerator and denominator polynomials of the transfer function,
respectively. Being the stability as a first, the function given in Eq. (\ref%
{5}) contains many important system information and concepts.

\section{STABILITY ANALYSIS OF FRACTIONAL DIFFERENTIAL EQUATIONS}

\setcounter{section}{3}

\noindent There are many ways of testing the stability of a linear
time-invariant differential equation. Stability of the differential equation
can be examined by applying Routh-Hurwitz test on its denominator
polynomial, checking the locations of the poles of its transfer function
whether they are on the left half s-plane or not and exploring its output
whether the output remains bounded with an impulse or step input excitation
\cite{35}. If the differential equation involves time-delay or fractional
order terms, in this case, Routh-Hurwitz' criteria cannot be applied.

\bigskip The denominator seen in Eq. (\ref{5}) of a FDE is in the form of a
quasi-polynomial and it is expressed by%
\begin{equation}
D(s)=a_{n}s^{a_{n}}+a_{n-1}s^{a_{n-1}}+\cdots +a_{1}s^{a_{1}}+a_{0}.  \tag{6}
\label{6}
\end{equation}%
For the stability analysis, the quasi-polynomial in Eq. (\ref{6}) is
transformed to the following commensurate order quasi-polynomial

\begin{equation}
D_{c}(s)=\prod_{i=0}^{n}(s^{\alpha }+\lambda
_{i})=\prod_{i=0}^{n}P(s^{\alpha }),  \tag{7}  \label{7}
\end{equation}%
where $\alpha $ is the least common multiple of $\alpha _{1},\alpha
_{2},\cdots ,\alpha _{n}.$ Stability condition of this fractional order
polynomial was given by Matignon \cite{36} in 1996 as follows%
\begin{equation}
|arg(-\lambda _{i})|>\alpha \frac{\pi }{2},~~~~\forall i=1,2,\cdots ,n,
\tag{8}  \label{8}
\end{equation}%
where $-\lambda _{i}$ $(i=1,2,\cdots ,n)$ are the roots of the
pseudo-polynomial $P(s^{\alpha })$ \cite[38]{37}. Matignon's stability
analysis is applicable for only linear FDEs whose coefficients are known and
invariant. For the FDEs changing their coefficients/parameters in an
interval, Chen et al. \cite{27} proposed a very effective method for the
stability analysis. However, to investigate the stability may be more useful
in the case that the parameters of a FDE changes between minus and plus
infinity. In this paper, a method is presented for this type of stability
analysis. The method is based on obtaining stability boundaries and it
contains a graphical presentation. The most important property of this
method is to construct a conformal mapping from s plane to parameter space
composed by unknown parameters of the FDE. Therefore, the method is called
as parametric stability analysis and based on the D-decomposition method
(see \cite{39} for more detail).

In the D-decomposition method, there are three stability boundaries of a
polynomial \cite[40]{39}. The first boundary belongs to a real pole which
changes its stability property when passing through origin and crossing the
opposite half of $s-$plane with the parameter changes. Therefore, this
boundary is called real root boundary. It is obtained by putting zero
instead of $s$ in Eq. (\ref{6}) with

\begin{equation}
D(s)|_{s=0}=0\Rightarrow a_{0}=0.  \tag{9}  \label{9}
\end{equation}

Second boundary is named by infinite root boundary because of the fact that
the boundary belongs to a pole which changes stability property at infinity
with the parameter changes. Infinite root boundary is determined by
equalizing the coefficient of the greatest order term to zero;%
\begin{equation}
a_{n}=0.  \tag{10}  \label{10}
\end{equation}%
In order to obtain the boundaries from Eqs. (\ref{9}) and (\ref{10}), the
coefficients $a_{0}$ and $a_{n}$ should contain unknown parameters.
Otherwise, these boundaries do not exist for the considered FDE.

The last boundary results a couple of complex poles passing to one half
plane from the other half plane over the imaginary axis of the $s-$plane
with the parameter changes. This is the main boundary which determines the
stability region of the FDE and is named by complex root boundary. To obtain
this boundary, $s$ in Eq. (\ref{6}) is replaced by $j\omega $ as follows;

\begin{equation}
D(jw)={a_{n}}(j\omega )^{a_{n}}+a_{n-1}(j\omega )^{a_{n-1}}+\cdots
+a_{1}(j\omega )^{a_{1}}+a_{0}.  \tag{11}  \label{11}
\end{equation}%
Using the expansion $j^{x}=\cos (0.5\pi {x})+j\sin (0.5\pi {x})$ in Eq. (\ref%
{11}) for the fractional order powers of the complex number j, we get

\begin{equation}
D(jw)=\sum_{i=1}^{n}\left\{ a_{i}[\cos (0.5\pi {\alpha _{i}})+j\sin (0.5\pi {%
\alpha _{i}})]\omega ^{\alpha _{i}}\right\} +a_{0}.  \tag{12}  \label{12}
\end{equation}%
By decomposing $D(jw)$ into real and imaginary parts, we obtain

\begin{equation}
D_{t}(w)=\sum_{i=1}^{n}a_{i}\cos (0.5\pi {\alpha _{i}})\omega ^{\alpha
_{i}}+a_{0},~~~~~~D_{i}(w)=\sum_{i=1}^{n}a_{i}\sin (0.5\pi {\alpha _{i}}%
)\omega ^{\alpha _{i}}.  \tag{13}  \label{13}
\end{equation}%
By equalizing $D_{t}(\omega )$ and $D_{i}(\omega )$ to zero separately, two
variable equations system whose variables are the unknown parameters of the
FDE are obtained. The complex root boundary is found by solving of this
system with respect to $\omega $ for $0<\omega <\infty .$ The stability
region of the FDE is enclosed by these three boundaries in the parameter
space. The main property of this region is that all parameters in this area
make the FDE stable.

\section{PARAMETRIC STABILITY ANALYSIS OF FRACTIONAL DIFFERENTIAL EQUATIONS}

\setcounter{section}{4}

In this section, we consider the FDEs which are often encountered in the
engineering systems have the following form:
\begin{equation}
a_{\text{ }0}D_{t}^{\alpha _{2}}y(t)+b_{\text{ }0}D_{t}^{\alpha
_{1}}y(t)+cy(t)=ku(t),  \tag{14}  \label{14}
\end{equation}%
where $u(t)$ and $y(t)$ are forcing and response functions, respectively; $%
a,b,c$ and $k$ are real coefficients, $\alpha _{1}$ and $\alpha _{2}$ are
the fractional order powers such that $0<\alpha _{1}<\alpha _{2}<2.$

As it is pointed out in the previous section, the transfer function of a FDE
encloses important stability information. By taking the Laplace transform of
the FDE in Eq. (\ref{14}), the transfer function giving input-output
relation is obtained as

\begin{equation}
G(s)=\frac{Y(s)}{U(s)}=\frac{k}{as^{\alpha _{2}}+bs^{\alpha _{1}}+c}.
\tag{15}  \label{15}
\end{equation}%
If the coefficients $a,b,c$ and the fractional orders $\alpha _{1}$ and $%
\alpha _{2}$ are constant or they change in a specific interval, the
stability of Eq. (\ref{15}) can be determined by Matignon's method \cite{36}
and Chen et al's method \cite{27} easily as mentioned in Section 3. Here, we
investigate the parametric stability for the full variation range of these
coefficients using the D-decomposition method.

For obtaining the stability boundaries, we consider the denominator of Eq. (%
\ref{15})

\begin{equation}
D(s)=as^{\alpha _{2}}+bs^{\alpha _{1}}+c.  \tag{16}  \label{16}
\end{equation}%
The real root boundary for Eq. (\ref{16}) is determined as

\begin{equation}
c=0  \tag{17}  \label{17}
\end{equation}%
Since $\alpha _{2}>\alpha _{1}$, the infinite root boundary is found by
applying Eq. (\ref{10}) to Eq. (\ref{16}) as the follows

\begin{equation}
a=0.  \tag{18}  \label{18}
\end{equation}%
For the complex root boundary, the real and imaginary parts of Eq. (\ref{13}%
) are equalized to zero and the following system of equations is obtained:

\begin{equation}
a\omega ^{\alpha _{2}}\cos (0.5\pi {\alpha _{2}})+b\omega ^{\alpha _{1}}\cos
(0.5\pi {\alpha _{1}})+c=0,  \tag{19}  \label{19}
\end{equation}

\begin{equation}
a\omega ^{\alpha _{2}}\sin (0.5\pi {\alpha _{2}})+b\omega ^{\alpha _{1}}\sin
(0.5\pi {\alpha _{1}})=0,  \tag{20}  \label{20}
\end{equation}%
By solving this system of equations with respect to $b,\alpha _{1},\alpha
_{2}$ for the coefficients $a$ and $c$, we get the following formulas:

\begin{equation}
a=-b\omega ^{{\alpha _{1}}-{\alpha _{2}}}\frac{\sin (0.5\pi {\alpha _{1}})}{%
\sin (0.5\pi {\alpha _{2}})}  \tag{21}  \label{21}
\end{equation}

\begin{equation}
c=-b\omega ^{{\alpha _{1}}}\frac{\sin \left[ 0.5\pi ({\alpha _{2}-\alpha _{1}%
})\right] }{\sin (0.5\pi {\alpha _{2}})}.  \tag{22}  \label{22}
\end{equation}%
By changing $\omega $ from zero to infinity for different values of $%
b,\alpha _{1}$ and $\alpha _{2}$, the complex root boundaries are obtained.

\textbf{Definition 4.1: }When three boundary conditions defined in Eqs. (\ref%
{17}), (\ref{18}), (\ref{21}) and (\ref{22}) are drawn in $(a,c)-$parameter
plane together, the plane separates to many regions. The most basic property
of these regions is that all points in any region have the same number of
stable or unstable roots \cite{39}. With this reference, the region whose
all poles are stable are called as stability region. Every point in this
region makes the FDE in Eq. (\ref{14}) stable. The stability of each region
can be determined by selecting a test point in the region and checking the
stability of Eq. (\ref{16}) in every region \cite{36}. Plot of stability
boundaries and stability regions on $(a,c)-$parameter plane will yield the
parametric investigation of the stability of the given fractional order
system. This provides for example to design a more robust system.

\section{SIMULATION EXAMPLES}

\setcounter{section}{5} \setcounter{equation}{0}

In this section, stability analysis of some fractional differential
equations commonly used in the literature are given for illustration of the
validity of the presented method. In the first example, fractional Basset
equation defining the dynamical motion of an object which submerged into a
fluid is considered. In the next example, as a more general case, stability
analysis is investigated for a commensurate order FDE. In the last example,
the stability ranges of the parameters $a,b$ and $c$ for an industrial
heating furnace is investigated.

\subsection{EXAMPLE 1}

\bigskip The motion dynamics of an object which is submerged into an
incompressible fluid is one of the frequently studied topics in engineering
literature. Basset \cite{41} proposed an equation and its solution for a
sphere moving in a viscous liquid when the sphere is moving in a straight
line under the action of a constant force, such as gravity, and also when
the sphere is surrounded by viscous liquid and is set in rotation about a
fixed diameter and then left to itself \cite{42}. This equation is named as
Basset equation and it is expressed by the following fractional order
differential equation%
\begin{equation}
a\text{ }_{0}D_{t}^{1}y(t)+b\text{ }_{0}D_{t}^{\alpha }y(t)+cy(t)=u(t),
\tag{23}  \label{23}
\end{equation}%
where $\alpha \in (0,1)$, and $a\neq 0,b,c$ are arbitrary real coefficients
\cite{43}. While the equation in (\ref{23}) is called by classical Basset
equation for $\alpha =0.5$, it is named as generalized Basset equation for $%
0<\alpha <1$ \cite{44}.

For the stability analysis of Basset equation, Govindaraj and Balachandran
proposed an analytical method in \cite{45}. Even though they found solutions
for the changes in the coefficients a,b and c separately, they have not
presented a general solution with respect to changing in the coefficients.
The goal in this example is to obtain a simple graphical result giving the
stability or instability of the equation according to the changing of the
parameters for the system given in Eq. (\ref{23}).

Real and infinite root boundaries are the same with in Eqs. (\ref{17}) and (%
\ref{18}), respectively. For the complex root boundary, putting $\alpha
_{2}=1$ and $0<\alpha _{1}=\alpha <1$ in Eqs. (\ref{21}) and (\ref{22}), the
expressions giving complex root boundary are found by

\begin{equation}
a=-b\omega ^{\alpha -1}\sin \alpha \frac{\pi }{2},  \tag{24}  \label{24}
\end{equation}

\begin{equation}
c=-b\omega ^{\alpha }\cos \alpha \frac{\pi }{2}.  \tag{25}  \label{25}
\end{equation}%
For the stability analysis of Eq. (\ref{23}), we consider the classical
Basset equation firstly. For $\alpha =0.5$, the stability boundaries are
\begin{equation}
\text{Real~root~boundary:}~~~c=0,  \tag{26}  \label{26}
\end{equation}%
\begin{equation}
\text{Infinite~root~boundary:}~~~a=0,  \tag{27}  \label{27}
\end{equation}%
\begin{equation}
\text{Complex~root~boundary:}~~~a=-\frac{\sqrt{2}}{2}b\omega ^{-0.5}~~\text{%
and~}~c=-\frac{\sqrt{2}}{2}b\omega ^{0.5}=0.  \tag{28}  \label{28}
\end{equation}

\begin{center}
\includegraphics[height=85mm]{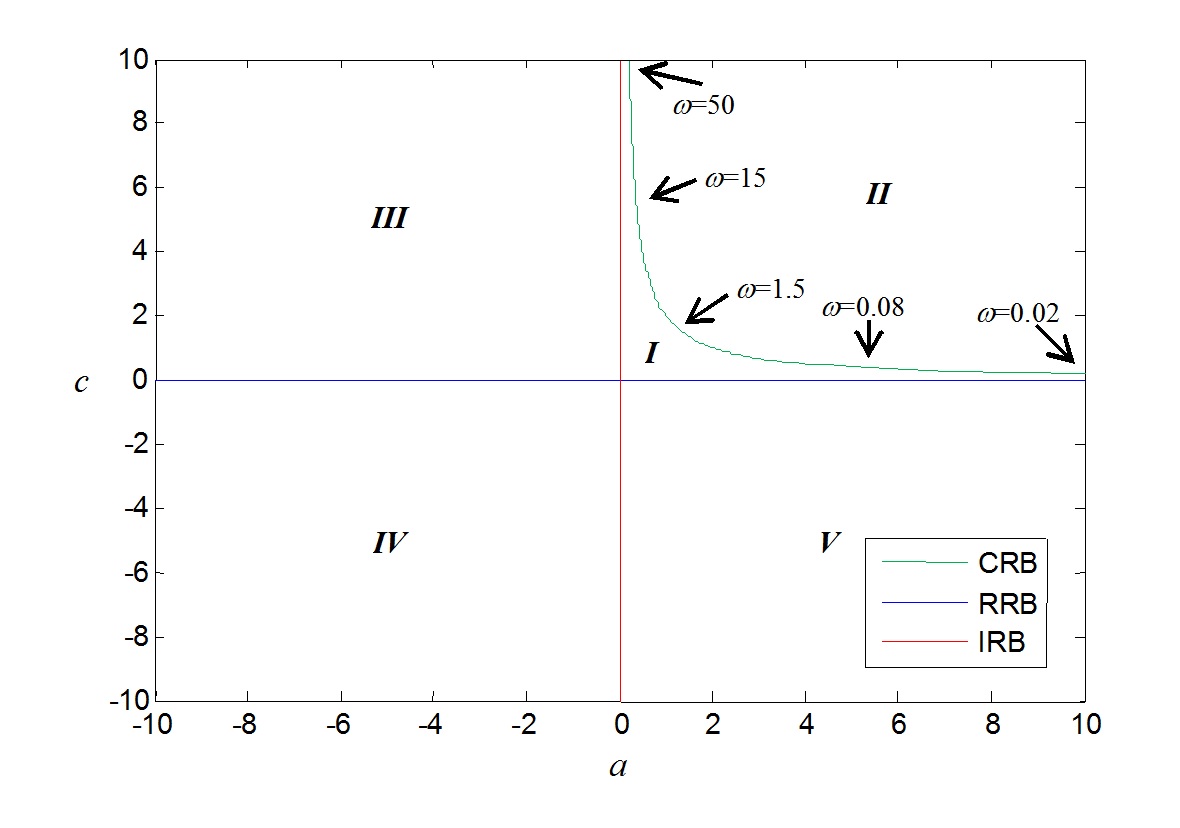} \ \ \ \

\textbf{Figure 1:} Stability boundaries of the classical Basset Equation for
$b=-2$\ \ \ \
\end{center}

For changing $\omega $ from $0$ to $\infty $, these boundaries decompose the
$(a,c)-$plane into many number of regions for various values of b. For
example, the stability boundaries constitute five regions in the $(a,c)-$%
plane for the value of $b=-2$ as shown in Fig. 1 where complex root boundary
is $ac=\frac{b^{2}}{2}=2,$ $a,c>0$ from Eq. (\ref{28}). Since the regions
are unlimited throughout the axes of $a$ and $c$, the figure is limited for
good visibility in the interval of $[-10,10]$ for these axes. The most
important characteristic of these regions is that all points in every region
have the same number of stable and unstable roots. Because of this reason,
to determine which regions are stable or not among these five areas, it is
sufficient to select only one testing point from every region and checking
the stability of Eq. (\ref{23}) according to these points. As shown in Fig.
2, it is found that the second and fourth regions are the stability regions.
For verification of these regions, it can be seen that the results are
suitable with the following results

\begin{equation*}
\text{(Asymptotically) stable for}~a=-3,b=-2,c=-4
\end{equation*}%
\begin{equation*}
\text{(Periodically) stable for}~a=1,b=-2,c=2
\end{equation*}%
\begin{equation*}
\text{Unstable for}~a=1,b=-2,c=-1
\end{equation*}

\begin{center}
\includegraphics[height=85mm]{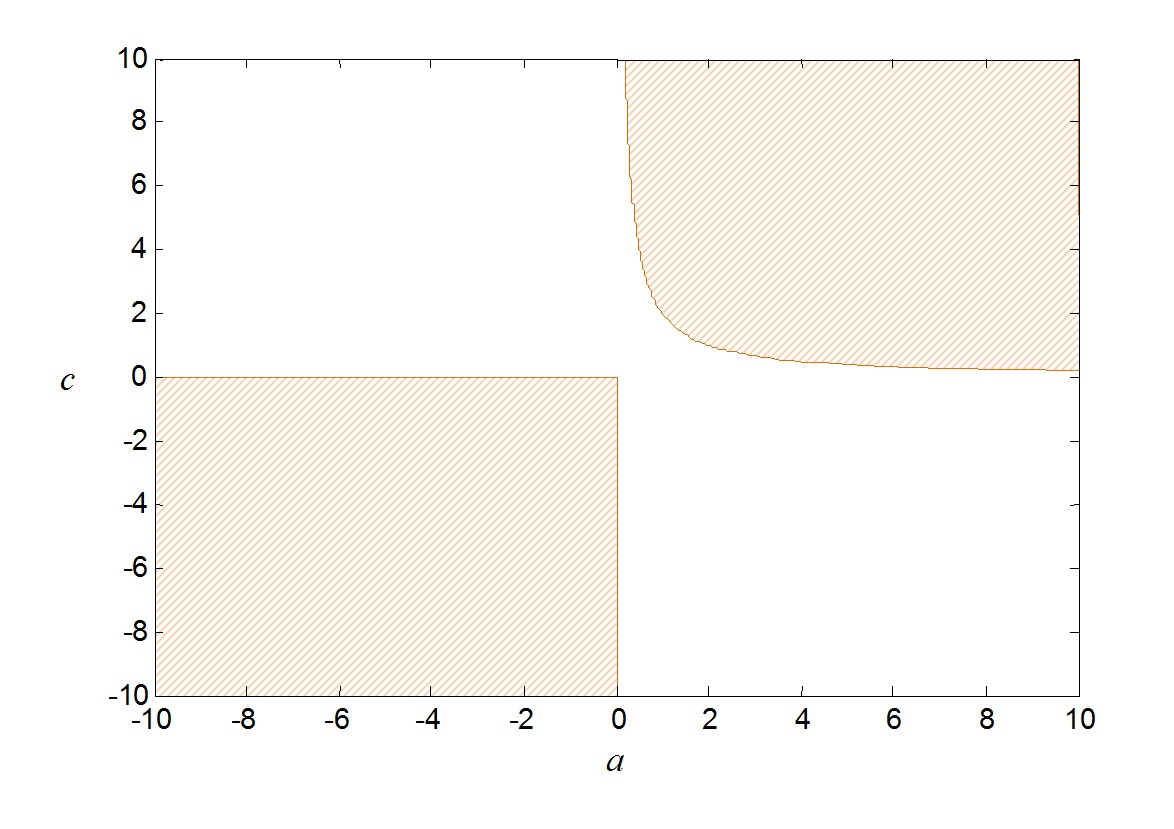} \ \ \ \

\textbf{Figure 2:} Stability region of the classical Basset Equation for $%
b=-2$\ \ \ \
\end{center}

\noindent which are given by Govindaraj and Balachandran \cite{45} for the
classical Basset equation. By varying b and repeating the above procedure,
different stability regions are obtained for each $b$ value as shown in Fig.
3. It is seen from this figure that smaller values of $|b|$ provide bigger
stability regions. To illustrate the graphical results more clearly, the
overall stability region can then be visualized in a 3-D plot as shown in
Fig. 4.

\begin{center}
\includegraphics[height=55mm]{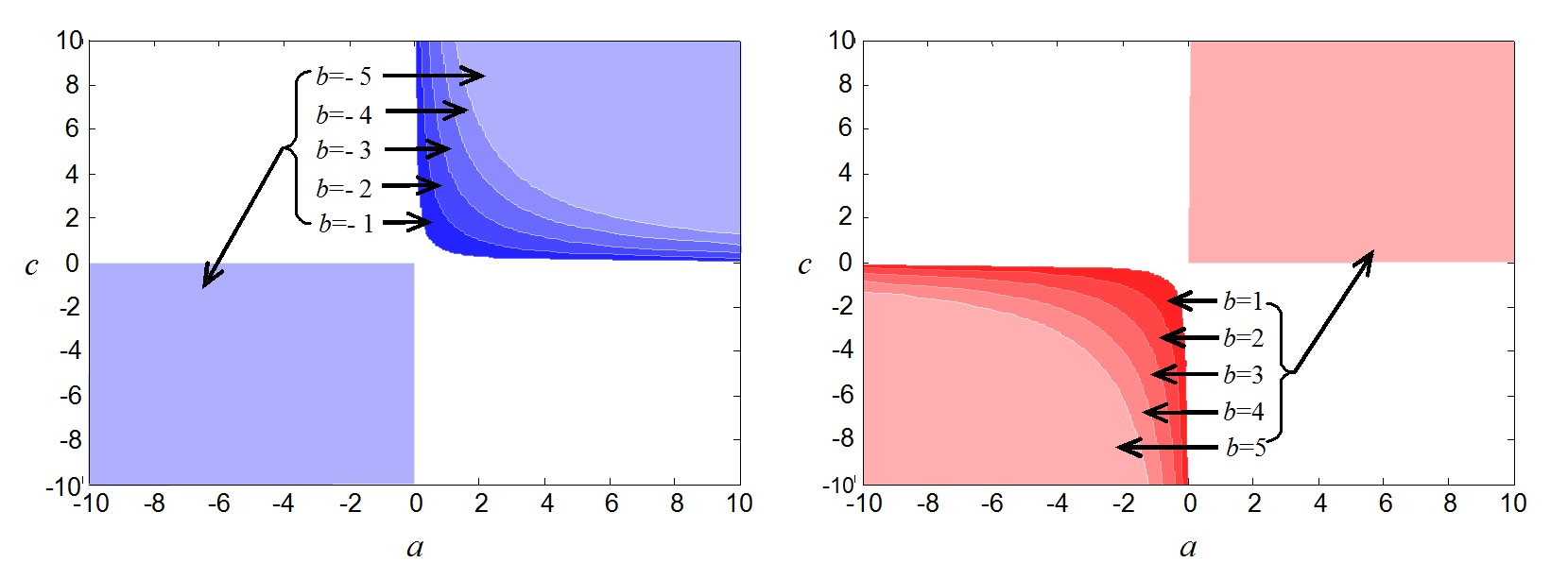} \ \ \ \

\textbf{Figure 3:} Changing the stability regions of the classical Basset
equation for various values of $b$

\includegraphics[height=85mm]{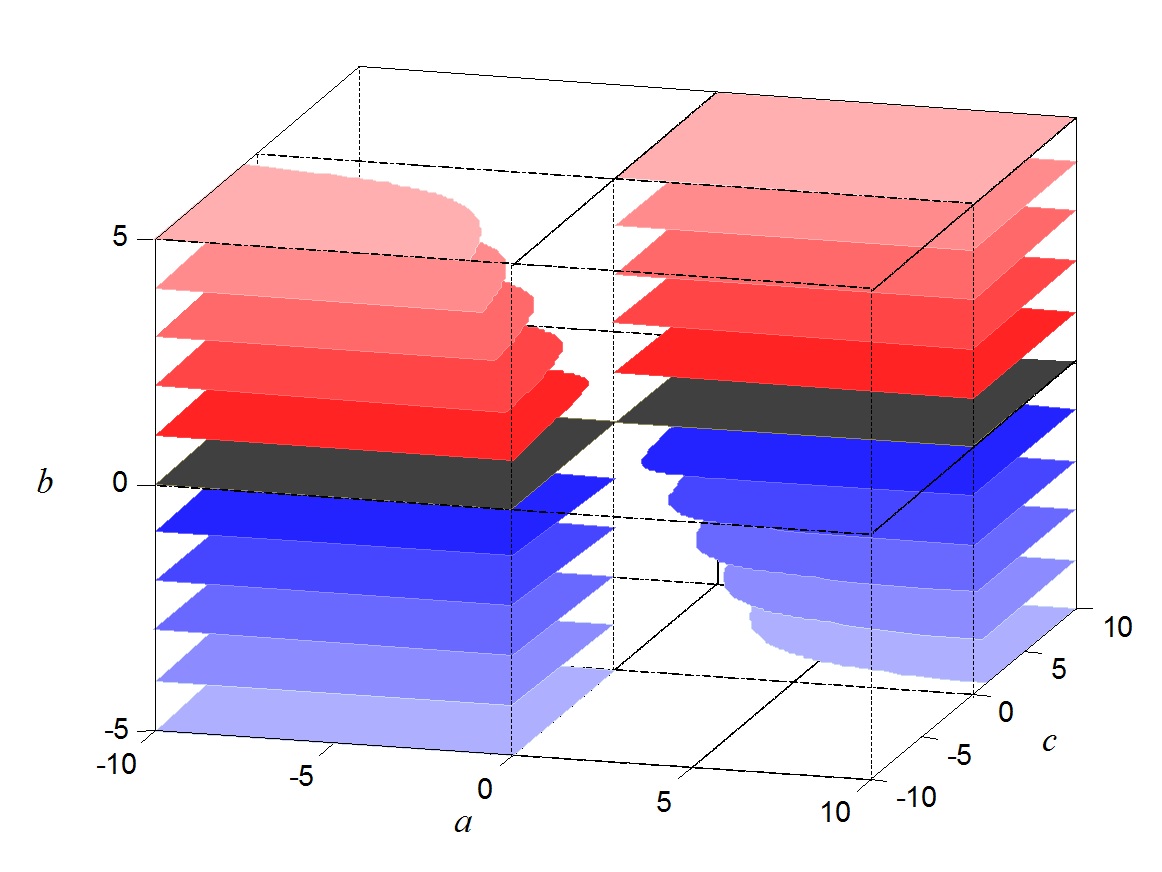} \ \ \ \

\textbf{Figure 4:} The overall 3-D stability region for the classical Basset
equation\ \ \ \
\end{center}

In order to make a more general stability analysis, we consider the
generalized Basset equation for $0<\alpha <1$ lastly. Fig. 5 shows the
stability regions of the generalized Basset equation for different values of
the parameter $\alpha $ for $b=-2$ and $b=4$. Notice that the change of $%
\alpha $ affects only the curve

\begin{center}
\includegraphics[height=62mm]{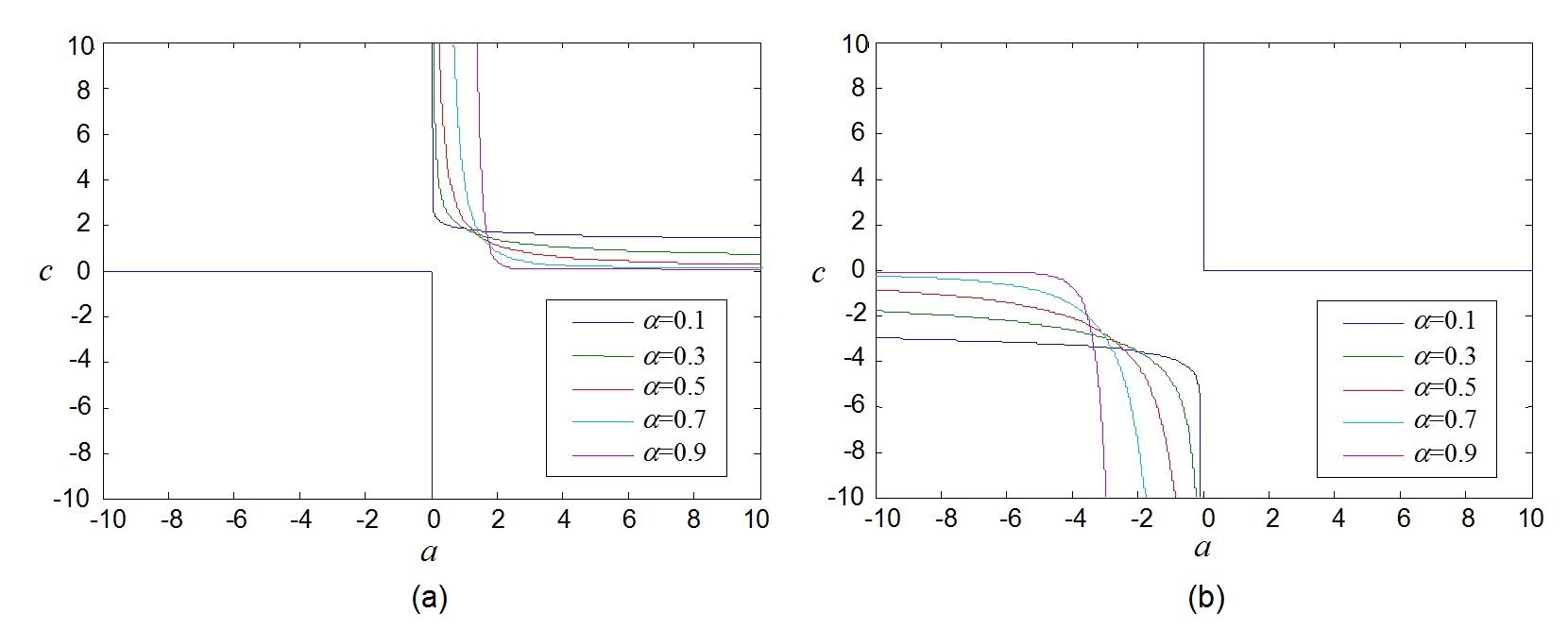} \ \ \ \

\textbf{Figure 5:} Stability regions of the generalized Basset equation for
different values of $\alpha ;$ a) $b=-2,$ b) $b=4$\ \ \

\includegraphics[height=85mm]{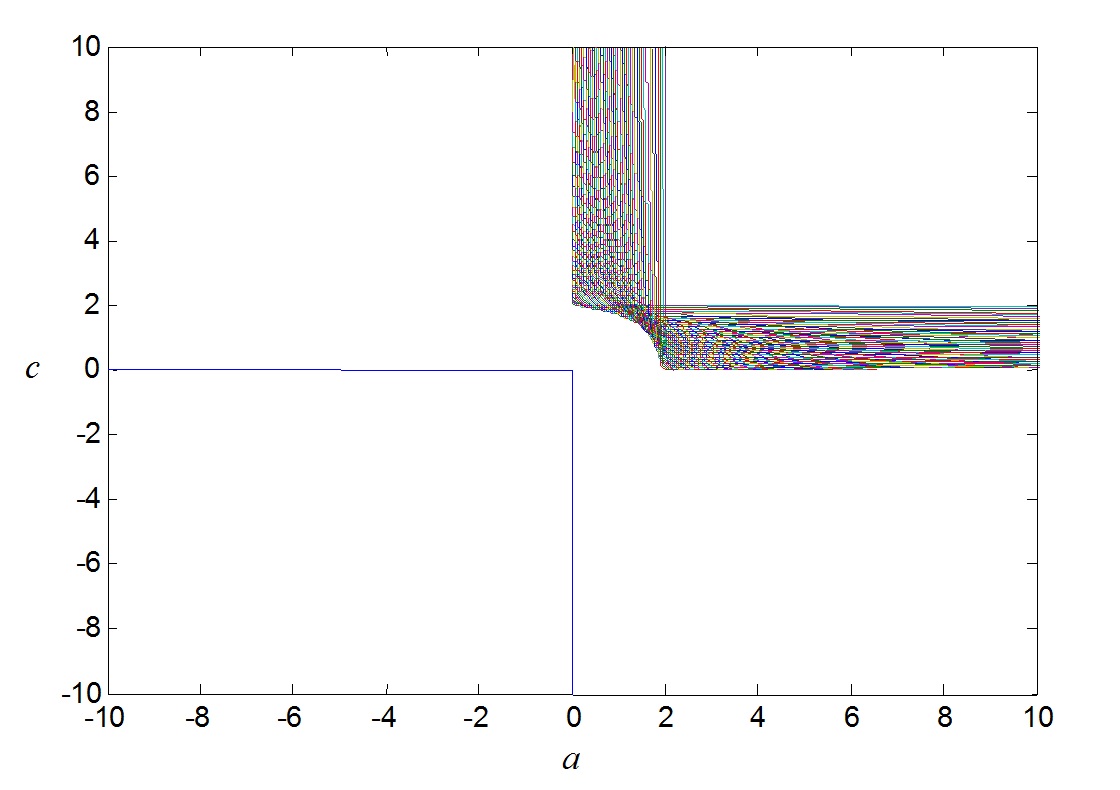} \ \ \ \

\textbf{Figure 6:} The variation of stability regions of generalized Basset
equation for $b=-2$ as $\alpha $ changes in $(0,1)$ with the increments of $%
0.01$\ \ \ \
\end{center}

\noindent of complex root boundary but does not influence the other
boundaries. However, the variation of the complex root boundary remains only
in a bounded area with the variation of $\alpha $. To generalize this fact
clearly, the change of complex root boundary is plotted for $100$ different
values of $\alpha $ in the interval of $(0,1)$ for $b=-2$ in Fig. 6. Robust
stability region can be defined for the area taking the shape of a rectangle
$-b=2<a,c\leq 10$ which remains inside the set of complex root boundaries
and also the rectangular area $-10\leq a,c<0$ at the left bottom. This
robust region always gives the guaranteed stable results for all values of $%
a,b,c$ without depending on the values of $\alpha $ in $(0,1)$. With
reference to this result, three dimensional robust stability region which
does not depend on the parameter $\alpha $ of generalized Basset equation is
shown in Fig. 7. Any $(a,b,c)$ point selected in this region makes the
fractional Basset equation in Eq. (\ref{23}) absolutely stable for any value
of $\alpha $ in $(0,1)$. One of the other advantages of obtaining three
dimensional robust stability region is that it gives the limit of how much
any parameter can be changed without affecting the stability of the Basset
equation for any point in the stability region. For example, from Fig. 7, it
is clear that the Basset equation is stable for $a=-5,$ $b=1$ and $c=-10$.

\begin{center}
\includegraphics[height=85mm]{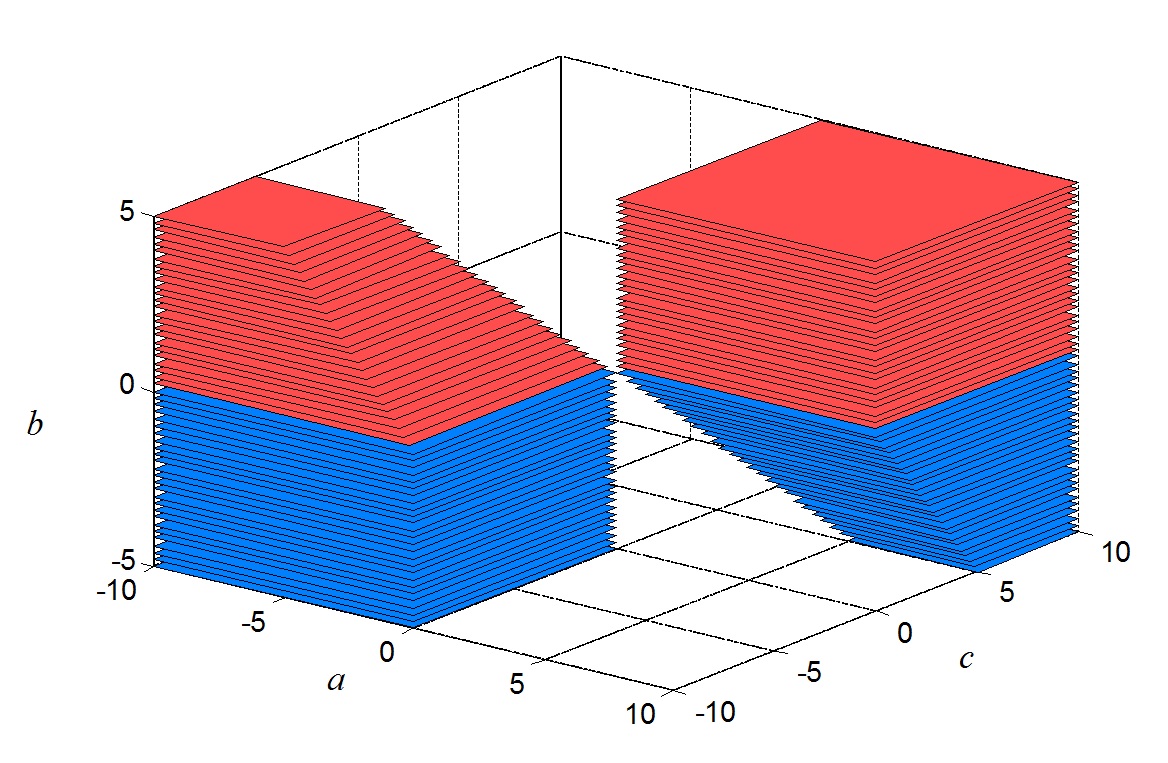} \ \ \ \

\textbf{Figure 7:} 3-D robust stability region independent of $\alpha $\
parameter of the generalized Basset equation\ \ \ \
\end{center}

\noindent As reference to this result, it is possible to say that the
stability of Basset equation is not affected by the change in the values of $%
a$ and $c$ parameters in the range of $(-\infty ,5)$ and $b$ parameter in
the range of $(-\infty ,5)$.

\textbf{Remark 5.1.} It can be observed from Fig. 1 that if the complex root
boundary curve which is derived for any $b$ value of commensurate order
equation obtained for $\alpha =0.5$ passes through any $(a,c)$ point, it
also passes from the point $(c,a)$ symmetrically. This result originates
from the powers of Eq. (\ref{14}) to be commensurate orders of $1$ and $0.5$%
. However, this characteristic is invalid for the general incommensurate
orders, i.e. when the value of $\alpha $ is different from $0.5$.

\subsection{EXAMPLE 2}

\bigskip Commensurate order FDEs are commonly used for modelling of physical
systems and industrial processes \cite{46}. In this example, the following
equation containing two fractional terms is considered:

\begin{equation}
a\text{ }_{0}D_{t}^{2\alpha }y(t)+b\text{ }_{0}D_{t}^{\alpha
}y(t)+cy(t)=u(t).  \tag{29}  \label{29}
\end{equation}%
This equation represents a FDEs family for different values of $\alpha $ and
it also contains classical Basset equation for $\alpha =0.5$. Members of
this family are named by multi-term differential equations if the power of
the greatest derivative term is greater than 1 (or $\alpha >0.5$) and
single-term differential equations if the power of the greatest derivative
term is less than $1$ (or $\alpha <0.5$) \cite{47}.

Expressions which belong to the complex root boundary of commensurate order
differential equation in Eq. (\ref{29}) whose real and infinite root
boundaries are given by Eqs. (\ref{17}) and (\ref{18}) are obtained for $%
\alpha _{2}=2\alpha $ and $\alpha _{1}=\alpha $ in Eqs. (\ref{21}) and (\ref%
{22}) as follows:%
\begin{equation}
a=-b\omega ^{-\alpha }[\sin (0.5\pi \alpha )/\sin (\pi \alpha )],  \tag{30}
\label{30}
\end{equation}%
\begin{equation}
c=-a\omega ^{2\alpha }\cos (\pi \alpha )-b\omega ^{\alpha }\cos (0.5\pi
\alpha ).  \tag{31}  \label{31}
\end{equation}%
For the various values of $\alpha $, the stability regions can be easily
obtained according to the values of $b$. For example, the stability regions
for $\alpha =0.2$ and $\alpha =0.8$ are plotted for $b\in \lbrack -5,-1]$ as
shown in Fig. 8 where the complex root boundary calculated from (\ref{30})
and (\ref{31}) is%
\begin{equation*}
ac=\frac{b^{2}}{2\left( 1+\cos \alpha \pi \right) };\text{ }a,c>0.
\end{equation*}
Stability regions for the values of $b$ changing in the interval $[1,5]$ are
replacements of the same

\begin{center}
\includegraphics[height=62mm]{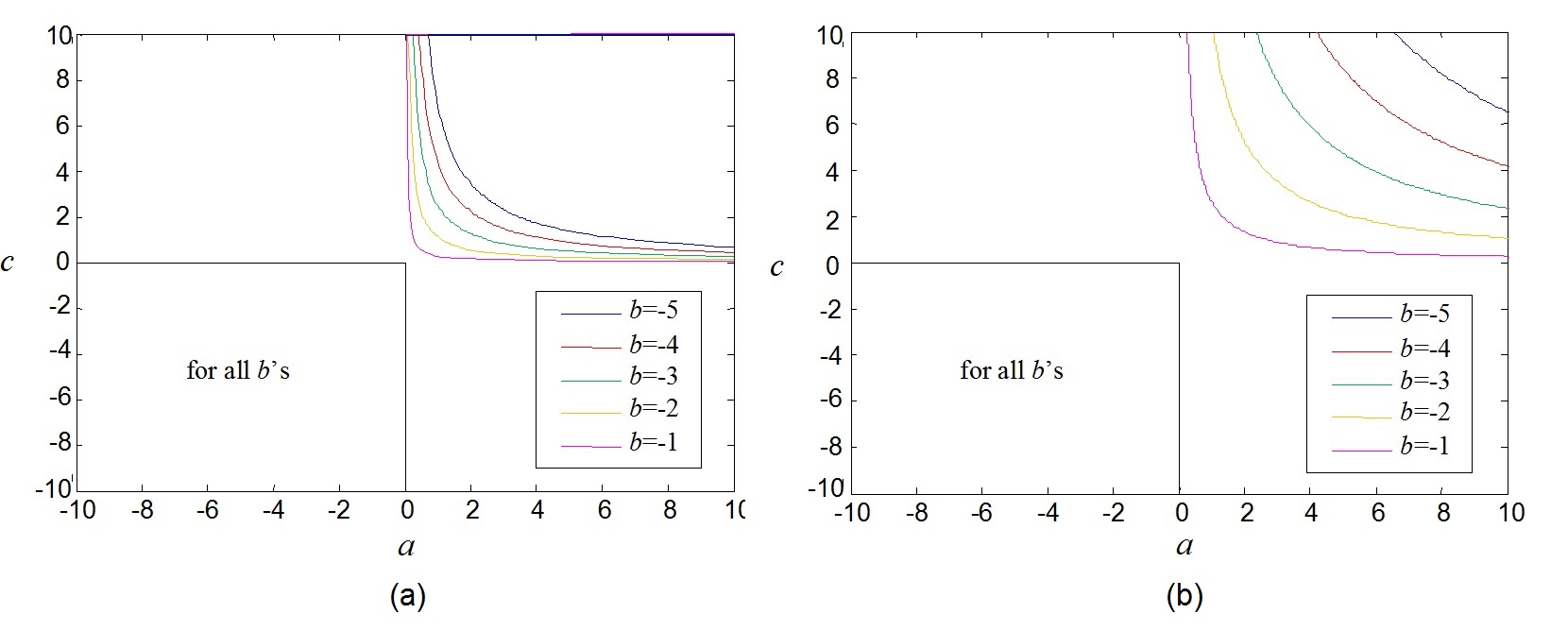} \ \ \ \

\textbf{Figure 8:} Stability regions for the values of the parameter $b$ in
the interval of $[-5,-1]$ as $\alpha $ changes$;$ a) $\alpha =0,2,$ b) $%
\alpha =0.8$\ \ \ \

\includegraphics[height=62mm]{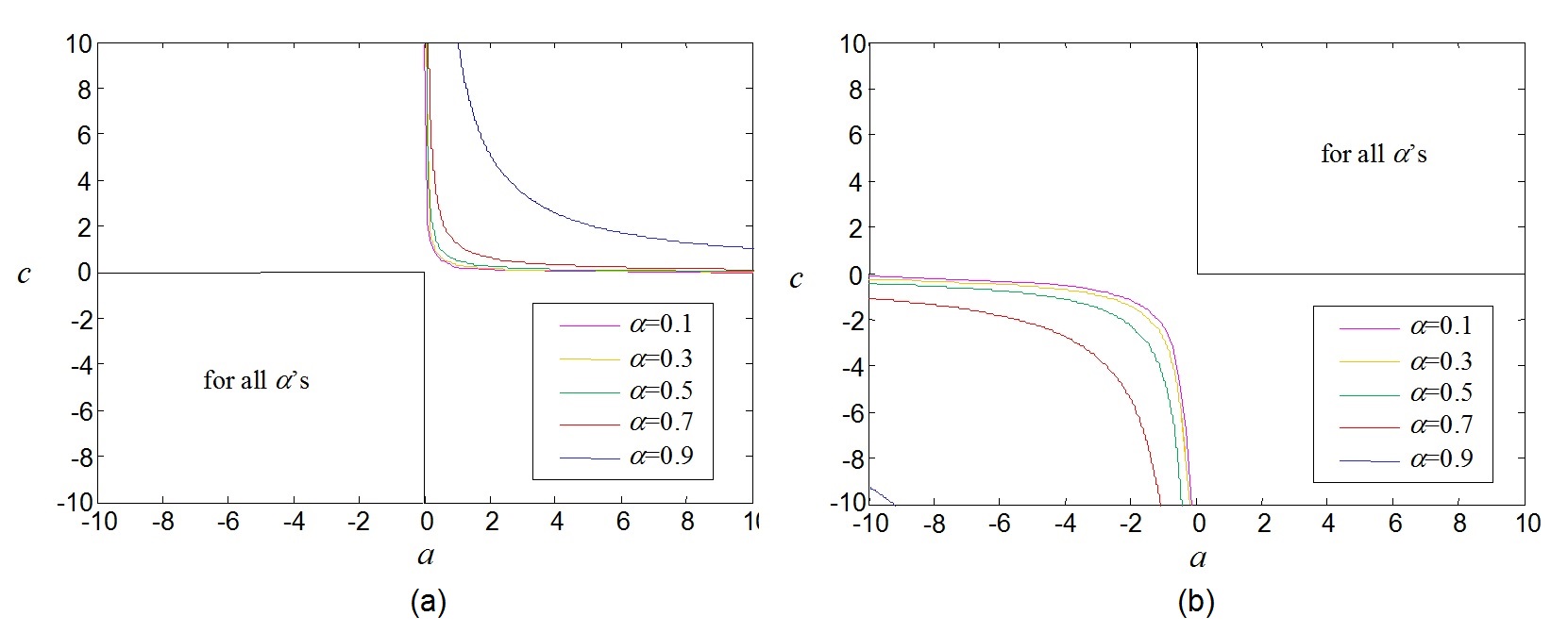} \ \ \ \

\textbf{Figure 9:} Stability regions for various values of $\alpha ;$ a) $%
b=-2,$ b) $b=3$\ \ \ \
\end{center}

\noindent regions with respect to origin symmetrically. In Fig. 9, the
stability regions for five different values of $\alpha $ are seen when $b=-2$
and $b=3$. For $b=-2$, when the values of $\alpha $ are changed more often
for instance, $100$ times in the range of $(0,1),$ the change of stability
region is appeared from Fig. 10. It is seen from this figure that the
stability region fills right upper part of the $(a,c)-$plane if the value of
$\alpha $ approaches to $0$ and the stability region is getting smaller if
the value of $\alpha $

\begin{center}
\includegraphics[height=85mm]{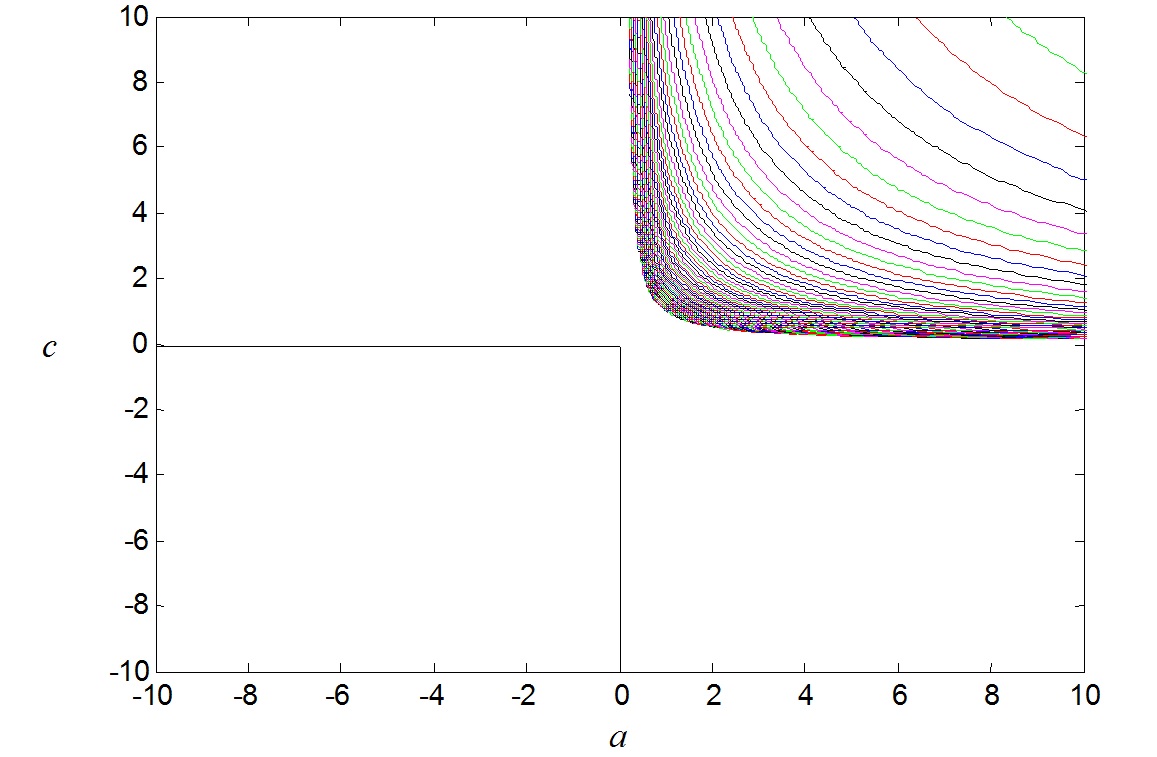} \ \ \ \

\textbf{Figure 10:} The variation of stability regions for $100$ values of $%
\alpha $ in $(0,1)$ when $b=-2$\ \
\end{center}

\noindent approaches to $1$. As a result, the stability of Eq. (\ref{29})
for any value of $\alpha $ depending on parameter varying can be analyzed
from the figures easily. However, it is seen that obtaining a robust
stability region is very difficult on the contrary of Example 5.1.

\subsection{EXAMPLE 3}

\bigskip In this example, the following incommensurate order FDE%
\begin{equation}
a_{\text{ }0}D_{t}^{1.31}y(t)+b\text{ }_{0}D_{t}^{0.97}y(t)+cy(t)=u(t)
\tag{32}  \label{32}
\end{equation}%
is considered for modelling an industrial heating furnace. In this equation,
nominal values of $a,b$ and $c$ parameters are given as $a=14994,b=6009.5$
and $c=1.69$ in \cite{48}. Sondhi and Hote \cite{49} has shown the stability
of the equation for these nominal values. The goal in this example is to
verify this stability results and to determine the stability intervals by
assuming these parameters are varying.

Stability boundaries for Eq. (\ref{32}) are as follows:

\begin{equation}
\text{Real~root~boundary:}~~~c=0,  \tag{33a}  \label{33a}
\end{equation}%
\begin{equation}
\text{Infinite~root~boundary:}~~~a=0,  \tag{33b}  \label{33b}
\end{equation}%
\begin{equation}
\text{Complex~root~boundary:}~~~a=-1.1303b\omega ^{-0.34}~~\text{and~}%
~c=-0.576b\omega ^{0.97}.  \tag{33c}  \label{33c}
\end{equation}%
The stability region of the system for $b=6009.5$ is shown in Fig. 11 where
from (\ref{33c}) the lower left stability region is found bounded by the
curve%
\begin{equation}
\left( -a\right) \left( -c\right) ^{0.350515464}=118190.408.  \tag{33d}
\label{33d}
\end{equation}%
It is seen from this figure, the nominal values of $a$ and $c$ stay in the
stability region. So, the stability result of Sondhi and Hote's has been
verified easily without any complicated calculations. For a more general
case, three-dimensional stability region derived for various values of $b$
is seen in Fig. 12. From this figure, evolution of the stability as contrast
to system parameter changes can be investigated. The stability is conserved
at all positive values of $a$ and $c$, and at all negative values of $a$ and
$c$ staying in the oval part when $b$ is positive; at all negative values of
$a$ and $c$ and at all positive values of $a$ and $c$ staying in the oval
part when $b$ is negative.

\begin{center}
\includegraphics[height=85mm]{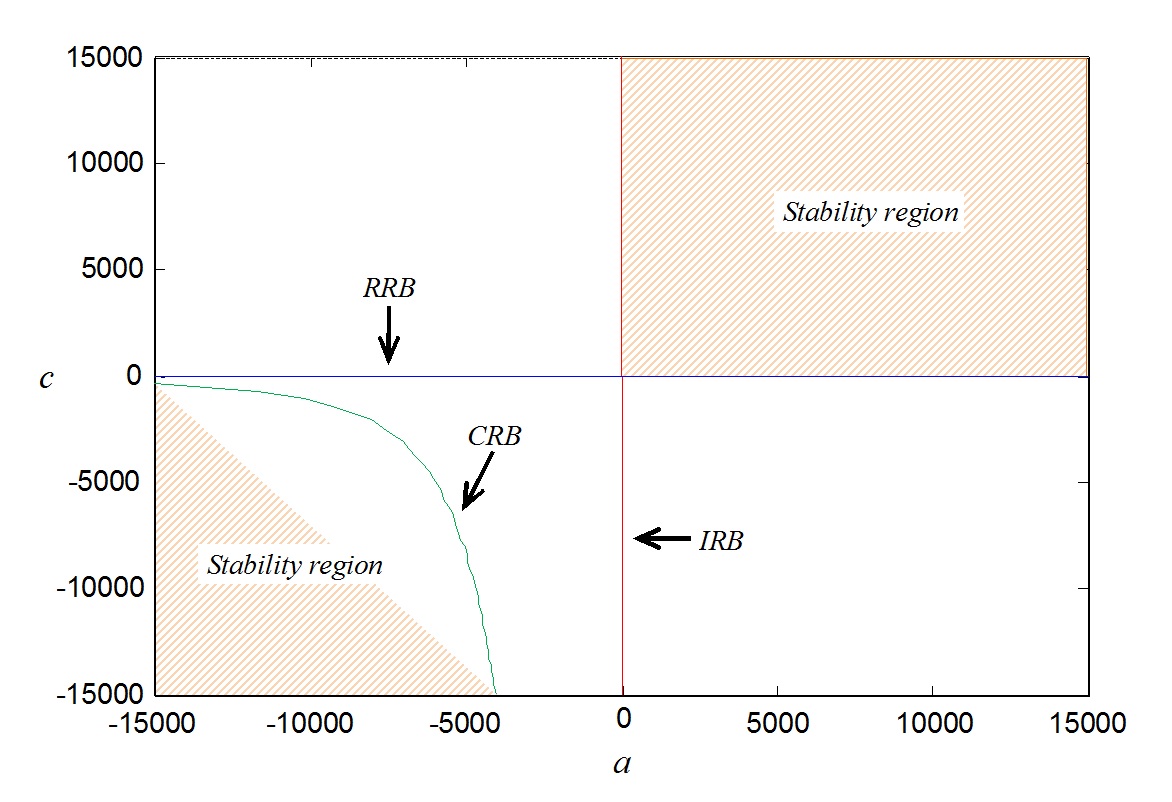} \ \ \ \

\textbf{Figure 11:} Stability region for industrial heating furnace \ \ \ \

\includegraphics[height=85mm]{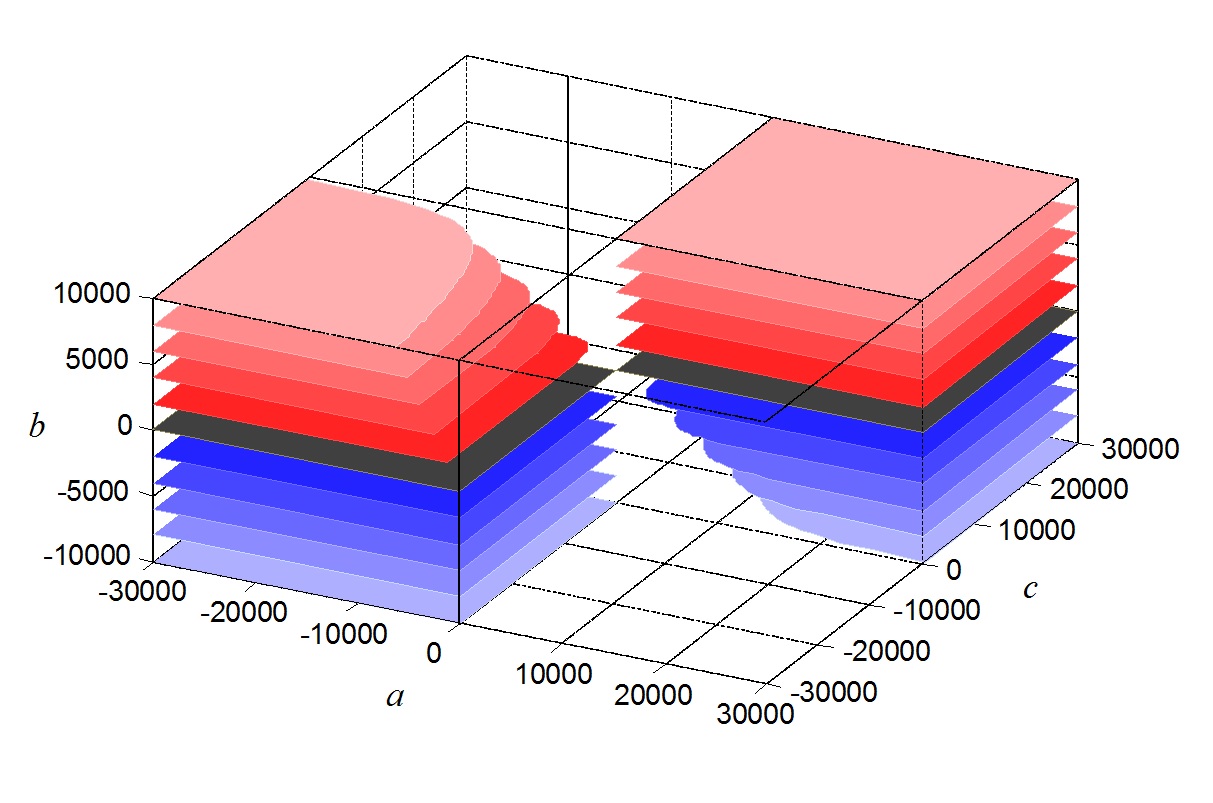} \ \ \ \

\textbf{Figure 12:} Three-dimensional overall stability region as $b$
changes in $[-10000,10000]$ \ \ \ \
\end{center}

\section{CONCLUSIONS}

\bigskip In this paper, a graphical based stability analysis method is
presented for the FDEs. The analysis concept is based on the derivation of
the stability boundaries and then the determination of the stability region
including the parameter set which makes the FDE stable. One of the most
important advantages of the method is that the stability analysis is done in
a visual environment without considering complex analytical solutions. This
method can be used not only for the investigation of the stability of a
differential equation whose parameters are not changed but also for
observation of the parametric robust stability of the equation whose
parameters are varying in a large interval. From this aspect, having a large
usage perspective of the method in comparison with the other methods is one
of the other advantages of the method. This provides opportunity to
engineers in their analyses about discussion more detail. Simulation
examples have been selected from the benchmark problems encountered in
engineering systems. As evidenced by the results given in these examples, it
can be concluded that the proposed graphical based method is reliable method
not only for stability analysis but also parametric robust stability
analysis according to the parameter changes.

The presented method can be generalized for stability analyses of fractional
differential equations having time delay which is a very popular subject in
the last decade. Moreover, the differential equations having more fractional
terms can be also studied. Here, when the number of unknown parameters
increases, the three dimensional graphs will be insufficient. In this case,
more than one graphs or four dimensional graphs with the fourth dimension
assuming by colour can be used.

\end{document}